\documentclass[prd,aps,twocolumn,showpacs, nofootinbib,superscriptaddress,notitlepage]{revtex4-1}
\usepackage{amsmath,graphicx,epsfig,amssymb}
\usepackage{inputenc}
\usepackage[usenames]{color}
\usepackage{slashed}
\usepackage{multirow}
\usepackage{subfigure}
\usepackage{dsfont}
\usepackage{comment}

\allowdisplaybreaks

\usepackage{floatrow}
\newfloatcommand{capbtabbox}{table}[][\FBwidth]

\begin{document}

\title{Improved  method to determine the   $\Xi_c-\Xi_c'$ mixing}

\author{Hang Liu}
\affiliation{INPAC, Key Laboratory for Particle Astrophysics and Cosmology (MOE),  Shanghai Key Laboratory for Particle Physics and Cosmology, School of Physics and Astronomy, Shanghai Jiao Tong University, Shanghai 200240, China}

\author{Wei Wang}
\email{Corresponding author: wei.wang@sjtu.edu.cn}
\affiliation{INPAC, Key Laboratory for Particle Astrophysics and Cosmology (MOE),  Shanghai Key Laboratory for Particle Physics and Cosmology, School of Physics and Astronomy, Shanghai Jiao Tong University, Shanghai 200240, China}
\affiliation{Southern Center for Nuclear-Science Theory (SCNT), Institute of Modern Physics, Chinese Academy of Sciences, Huizhou 516000, Guangdong Province, China}

\author{Qi-An Zhang}
\email{Corresponding author: zhangqa@buaa.edu.cn}
\affiliation{School of Physics, Beihang University, Beijing 102206, China}

\begin{abstract}
We develop an improved method to explore the  $\Xi_c- \Xi_c'$ mixing which arises from the flavor SU(3) and heavy quark symmetry breaking.  In this method, the flavor eigenstates under the SU(3) symmetry are at first constructed and the corresponding masses can be nonperturbatively determined. Matrix elements of the mass operators which break the flavor SU(3) symmetry sandwiched by the flavor eigenstates are then calculated. Diagonalizing the corresponding matrix of Hamiltonian gives the mass eigenstates of the full Hamiltonian and determines the mixing. Following the previous lattice QCD calculation of $\Xi_c$ and $\Xi_c'$, and estimating an off diagonal matrix element, we extract the mixing angle between the $\Xi_c$ and $\Xi_c'$. Preliminary numerical results for the mixing angle confirm the previous observation that such mixing is incapable to explain the large SU(3) symmetry breaking in semileptonic decays of charmed baryons. 
\end{abstract}

\maketitle

\section{Introduction}

Remarkably recent experimental measurements of decay widths of semileptonic charmed baryon decays have revealed a significant breakdown of flavor SU(3) symmetry~\cite{BESIII:2015ysy, BESIII:2023vfi, Belle:2021crz, Belle:2021dgc}, a pivotal tool extensively employed for deciphering weak decays of heavy mesons (for some recent applications, please see Refs.~\cite{Lu:2016ogy,He:2018php,He:2018joe,Wang:2018utj}). This pattern is in contradiction with the data on heavy bottom meson and baryon decays~\cite{ParticleDataGroup:2022pth} which to a good accuracy respects the flavor SU(3) symmetry.  In the pursuit of understanding this phenomenon, mechanisms were explored in the work~\cite{He:2021qnc}, with a very compelling contender being the incorporation of $\Xi_c-\Xi_c'$ mixing~\cite{Geng:2022yxb}. Actually, the mixing has been previously explored within
various models~\cite{Franklin:1981rc, Franklin:1996ve, Ito:1996mr, Aliev:2010ra, Matsui:2020wcc}. Subsequently,  very interesting works~\cite{Geng:2022yxb,Geng:2022xfz,Liu:2022igi,Ke:2022gxm} have explored the impact from $\Xi_c-\Xi_c'$ mixing in weak decays of charmed and doubly charmed baryons, and some interesting phenomena were discussed~\cite{Xing:2022phq}.

In a recent analysis to determine $\Xi_c-\Xi_c'$ mixing~\cite{Liu:2023feb}, four kinds of two-point correlation functions constructed by two kinds of baryonic operators are calculated using the technique of lattice QCD. Via the lattice data, two distinct methods are employed to extract the $\Xi_c-\Xi_c'$ mixing angle which is determined as $\theta=(1.2\pm0.1)^{\circ}$.  This small value is consistent with a previous lattice investigation in  Ref.~\cite{Brown:2014ena}, and determinations using QCD sum rules~\cite{Aliev:2010ra,Sun:2023noo}. 

In this work, we will not concentrate on the inconsistency in the angles obtained from the nonpertubative determination and the global fit. Instead, we focus on one ambiguity in defining the mixing angle between $\Xi_c$ and $\Xi_c'$ in the lattice simulation, which is equivalent to the construction of flavor SU(3) eigenstates in the simulation. Previous lattice QCD determination~\cite{Liu:2023feb} made use of the two-point correlation functions, in which an ambiguity exists in choosing the interpolating operators and accordingly in the extraction of the mixing angle.  In this work, we will develop an improved method to explore the  $\Xi_c- \Xi_c'$ mixing. 
In this method, the flavor eigenstates under the SU(3)symmetry are constructed at first and the corresponding masses are nonperturbatively determined. Three-point correlation functions made of the mass operator that breaks the SU(3) symmetry and the interpolating operators are then calculated. Taking a ratio with respect to the two-point correlation function removes the dependence in the interpolating operators and diagonalizing the corresponding matrix of  Hamiltonian unambiguously gives the mass eigenstates of the full Hamiltonian and determines the corresponding mixing. Using an off diagonal matrix element, we extract the mixing angle between the $\Xi_c$ and $\Xi_c'$. Though a sign ambiguity is left,  preliminary numerical results for the mixing angle confirm the previous observation that such mixing is incapable to explain the large SU(3) symmetry breaking in semileptonic charmed baryon decays. This leaves the problem of large SU(3) symmetry breaking observed in charmed baryon decays unresolved. 

The rest of this paper is organized as follows. In Sec.~\ref{sec:formalism}, we will give the theoretical formalism and the numerical results are collected in Sec.~\ref{sec:numerics}. We summarize this work in the last section. 

\section{Theoretical Formalism}
\label{sec:formalism}

\subsection{$\Xi_c$ and $\Xi_c'$ in SU(3) symmetry and mixing}

In the QCD Lagrangian for light quarks
\begin{eqnarray}
  \mathcal{L} & = & \bar{\psi} ( i \slashed{D} - M) \psi
\end{eqnarray}
with $D_\mu$ being the covariant derivative and 
\begin{eqnarray}
\psi =& \left(\begin{array}{c} u \\ d \\ s \end{array}\right),~~~
M =& \left(\begin{array}{ccc}
    m_u & 0 & 0\\
    0& m_d &0\\
    0 & 0 & m_s
  \end{array}\right),
\end{eqnarray}
the masses of three quarks are different and explicitly break the flavor SU(3) symmetry. In this work, we assume the isospin symmetry and adopt  $m_u = m_d \neq m_s$; consequently, $\mathcal{L}$ can be divided into two parts: the $SU(3)_F$ symmetry conserving term $\mathcal{L}_0$ and breaking term $\Delta \mathcal{L}$,
\begin{align}
\mathcal{L}_0 &=  \bar{\psi} ( i \slashed{D} - m_0) \psi ,~~~
\Delta  \mathcal{L}  =  -m_8 O_8,
\end{align}
with 
\begin{align}
m_0 &= \frac{2m_u+m_s}{3}, ~~~ m_8 = \frac{m_u-m_s}{\sqrt3}, \\
O_8 &= \frac{\bar uu + \bar dd - 2\bar ss}{\sqrt 3}. 
\end{align}

Therefore, the Hamiltonian can be derived as
\begin{align}
 H &= \int d^3 \vec{x}  \left[\frac{\partial\mathcal{L}}{\partial\dot{\psi}(\vec{x})}\dot{\psi}(\vec{x})+\frac{\partial\mathcal{L}}{\partial\dot{\bar{\psi}}(\vec{x})}\dot{\bar{\psi}}(\vec{x})-\mathcal{L}\right] \nonumber\\
 &\equiv H_0 + \Delta H,  
\end{align}
with 
\begin{eqnarray}
\Delta H= m_8  \int d^3 \vec{x}  O_8(\vec{x}). 
\label{eq:Hamiltonian_I}
\end{eqnarray}
In the Appendix, we also give an equivalent decomposition form as above.
 
In the heavy quark limit with $m_c\to \infty$, the heavy quark decouples from the light quark system. The interpolating operator for  a $J^P = (1 / 2)^+$
$\mathit{usc}$-type baryon  can be defined as
\begin{eqnarray}
  O & = & \epsilon^{abc} (q^{T a} C \Gamma s^{ b}) \Gamma' P_+
  \tilde{c}^c,
\end{eqnarray}
where $\tilde{c}$ denotes the heavy quark field in heavy quark effective theory (HQET) satisfying $\gamma^0
\tilde{c} = \tilde{c}$.  $P_+ = (1 + \gamma^0) / 2$ is the positive parity
projector. The totally antisymmetric tensor $\epsilon^{abc}$ is used
to sum over all color indices and guarantee the antisymmetric color
wavefunction. The transposition $T$ acts on a Dirac spinor, and $C = \gamma^0
\gamma^2$ is the charge conjugation matrix. The Dirac matrix $\Gamma$ and
$\Gamma'$ are related to the internal spin structures of the heavy baryon.

Neglecting $\Delta H$, the heavy baryon can be classified according to the flavor $SU(3)_F$ symmetry as $3 \otimes 3 = \bar{3} \oplus 6$, in which $\bar{3}$ denotes the
antisymmetric of light quark pair and its angular momentum is $J_{q s} =
0$, and $6$ denotes the symmetric case with $J_{q s} = 1$. Then the interpolating operators for the $J^P = (1 / 2)^+$
$\mathit{usc}$-type baryon can be chosen as~\cite{Grozin:1992td},
\begin{eqnarray}
  O_{SU(3)}^{\bar{3}} & = & \epsilon^{abc} (q^{T a} C
  \gamma_5 s^{ b}) P_+ \tilde{c}^c\\
  O_{SU(3)}^6 & = & \epsilon^{abc} (q^{T a} C \vec{\gamma}
  s^{b}) \cdot \vec{\gamma} \gamma_5 P_+ \tilde{c}^c .
\end{eqnarray}

These operators unambiguously define the corresponding flavor eigenstates $|\Xi^{\bar{3}}_c\rangle$ and $|\Xi^{6}_c\rangle$, which also act as the eigenstates of $H_0$,
\begin{eqnarray}
  H_0 | \Xi_c^{\bar{3} / 6}(\vec p=0) \rangle & = & m_{\Xi_c^{\bar{3} / 6}} |
  \Xi_c^{\bar{3} / 6}(\vec p=0) \rangle, 
\end{eqnarray}
where  $m_{\Xi_c^{\bar{3} / 6}}$ are the  mass eigenvalues in the case $\vec p=0$. 

When adding the $SU(3)_F$ breaking term
$\Delta H$, the mixing between $\Xi_c$ and $\Xi_c'$ will emerge (actually in the charmed baryon system, generating the  $\Xi_c-\Xi_c'$ mixing also requests to break the heavy quark symmetry).  One can easily
see that the $SU(3)_F$ breaking effect is characterized by $\Delta m = m_s - m_{u}$.   This  breaking effect leads to the mismatch between the flavor eigenstates and mass eigenstates
\begin{eqnarray}
    \left(
    \begin{array}{c}
       \left|\Xi_c\right\rangle \\ \left|\Xi_c'\right\rangle 
    \end{array}
    \right) = 
    \left(
    \begin{array}{cc}
       \cos\theta & \sin\theta \\ -\sin\theta & \cos\theta
    \end{array}
    \right)
    \left(
    \begin{array}{c}
       |\Xi^{\bar{3}}_c\rangle \\
       |\Xi^{6}_c\rangle 
    \end{array}
    \right), 
\end{eqnarray}
and in reverse, one has
\begin{eqnarray}
  \left(\begin{array}{c}
    | \Xi_c^{\bar{3}} \rangle\\
    | \Xi_c^6 \rangle
  \end{array}\right) & = & \left(\begin{array}{cc}
    \cos \theta & - \sin \theta\\
    \sin \theta & \cos \theta
  \end{array}\right) \left(\begin{array}{c}
    | \Xi_c \rangle\\
    | \Xi_c' \rangle
  \end{array}\right), 
\end{eqnarray}
where $\theta$ is the mixing angle, and the mass eigenstates are orthogonal,
\begin{align}
    H\left|\Xi_c\right\rangle = m_{\Xi_c}\left|\Xi_c\right\rangle, \quad 
    H\left|\Xi_c'\right\rangle = m_{\Xi_c'}\left|\Xi_c'\right\rangle,
\end{align}
$m_{\Xi_c}$ and $m_{\Xi_c'}$ denote the physical baryon masses. 

\subsection{Determination of the mixing angle}

In the following we will give the method to extract the mixing through the calculation of Hamiltonian's matrix elements. 
Let us start from the spin-averaged matrix of mass eigenstates
\begin{eqnarray}
  && M_E (\vec{p}) \equiv \int \frac{d^3 \vec{p'}}{(2 \pi)^3}  \notag\\
    &  &\;\;\;\;\; \times
  \left(\begin{array}{cc}
    \langle \Xi_c (\vec{p}) | H | \Xi_c (\vec{p'}) \rangle & \langle \Xi_c
    (\vec{p}) | H | \Xi_c' (\vec{p'}) \rangle\\
    \langle \Xi_c' (\vec{p}) | H | \Xi_c (\vec{p'}) \rangle & \langle \Xi_c'
    (\vec{p}) | H | \Xi_c' (\vec{p'}) \rangle
  \end{array}\right). 
\end{eqnarray}

Since the $\Xi_c$ and $\Xi_c'$ are the eigenstates of the full Hamiltonian, the above matrix is diagonal. In particular, if  $\vec{p} = 0$, $E_{\vec{p}}^2 = m^2$, one has 
\begin{eqnarray}
  M_E (\vec{p} = 0) & \equiv & \left(\begin{array}{cc}
    2 m_{\Xi_c}^2 & 0\\
    0 & 2 m_{\Xi_c'}^2
  \end{array}\right) .
\end{eqnarray}

When one rotates the external states from energy eigenstates to $SU(3)_F$ flavor eigenstates, the nondiagonal terms will be nonzero due to the
mixing effect
\begin{widetext}
\begin{eqnarray}
   M_F (\vec{p})&\equiv & \int \frac{d^3 \vec{p'}}{(2 \pi)^3} 
  \left(\begin{array}{cc}
    \langle \Xi_c^{\bar{3}} (\vec{p}) | H | \Xi_c^{\bar{3}} (\vec{p'}) \rangle
    & \langle \Xi_c^{\bar{3}} (\vec{p}) | H | \Xi_c^6 (\vec{p'}) \rangle\\
    \langle \Xi_c^6 (\vec{p}) | H | \Xi_c^{\bar{3}} (\vec{p'}) \rangle 
    &\langle \Xi_c^6 (\vec{p}) | H | \Xi_c^6 (\vec{p'}) \rangle
  \end{array}\right)\notag\\
  & = & \int \frac{d^3 \vec{p'}}{(2 \pi)^3}  \left(\begin{array}{cc}
    \langle \Xi_c^{\bar{3}} (\vec{p}) | (H_0 + \Delta H) | \Xi_c^{\bar{3}}
    (\vec{p'}) \rangle & \langle \Xi_c^{\bar{3}} (\vec{p}) | \Delta H | \Xi_c^6
    (\vec{p'}) \rangle\\
    \langle \Xi_c^6 (\vec{p}) | \Delta H | \Xi_c^{\bar{3}} (\vec{p'}) \rangle 
    &\langle \Xi_c^6 (\vec{p}) | (H_0 + \Delta H) | \Xi_c^6 (\vec{p'}) \rangle
  \end{array}\right). 
\end{eqnarray}
\end{widetext}
The contributions  from  $H_0$ vanish in the nondiagonal terms due to
the orthogonality between $| \Xi_c^{\bar{3}} \rangle$ and $| \Xi_c^6 \rangle$. When considering the conservation of momentum and the external states are rest ($\vec p=0$), the above matrix can be reduced to
\begin{eqnarray} \label{eq:mixing_matrix}
M_F (\vec{p}=0)&=& \left(\begin{array}{cc}
    2m_{\Xi_c^{\bar{3} }}^2 & 0\\
    0 & 2m_{\Xi_c^{6 }}^2
  \end{array}\right)\nonumber\\
  &  +& m_8 \left(\begin{array}{cc} 
    \langle \Xi_c^{\bar{3}}  |  O_8   | \Xi_c^{\bar{3}}
      \rangle & \langle \Xi_c^{\bar{3}}   |   O_8  | \Xi_c^6 \rangle\\
    \langle \Xi_c^6   |    O_8   | \Xi_c^{\bar{3}}  ) \rangle 
    &\langle \Xi_c^6   |    O_8 | \Xi_c^6   \rangle
  \end{array}\right),
\end{eqnarray}
where we have omitted the momentum  in external states $\Xi_c^{\bar 3} (\vec p=0)$ and  $\Xi_c^6 (\vec p=0)$ and the space coordinate in the scalar operator $O_8(\vec{x}= 0)$.  

It is necessary to point out that all the elements of the above matrix can be calculated using nonperturbative tools like lattice QCD. The off diagonal term should be equal and in total there are five quantities (including two masses and three independent matrix elements within Eq.(\ref{eq:mixing_matrix})) to be calculated. Diagoanlizing this matrix provides us with a straightforward way to extract the mixing angle. 

Interestingly,  physical masses can be experimentally measured or numerically determined from lattice QCD. In this case,  one can actually determine the mixing angle by only calculating the off diagonal matrix elements. To show this feasibility,  one can perform a rotation from the mass eigenstates basis to the flavor eigenstates basis and obtain the relations between the elements of matrix $M_F$:
\begin{eqnarray}
  M_{F, 11} & = & 2 \cos^2 \theta m_{\Xi_c}^2 + 2 \sin^2 \theta m_{\Xi_c'}^2,\notag\\
  M_{F, 12} & = & 2 \cos \theta \sin \theta (m_{\Xi_c}^2 - m_{\Xi_c'}^2),\notag\\
  M_{F, 21} & = & 2 \cos \theta \sin \theta (m_{\Xi_c}^2 - m_{\Xi_c'}^2), \notag\\
  M_{F, 22} & = & 2 \sin^2 \theta m_{\Xi_c}^2 + 2 \cos^2 \theta m_{\Xi_c'}^2,
\label{eq:flavor-basis}
\end{eqnarray}
where only the $\vec p=0$ case is considered. 
Therefore, one can establish a relation between the correlation functions and Eq.~\eqref{eq:flavor-basis}: 
\begin{eqnarray}
  M_{F, 11} & = &2 m_{\Xi_c^{\bar{3}}}^2 + m_8 M_{8}^{\bar{3} - \bar{3}} \notag\\
  & = &  2
  \cos^2 \theta m_{\Xi_c}^2 + 2 \sin^2 \theta m_{\Xi_c'}^2,\notag\\
  M_{F, 22} &= &2 m_{\Xi_c^6}^2 + m_8 M_{8}^{6 - 6} \notag\\
  & = & 2 \sin^2 \theta
  m_{\Xi_c}^2 + 2 \cos^2 \theta m_{\Xi_c'}^2,\notag\\
  M_{F, 12} & =& m_8 M_{8}^{\bar{3} - 6} \notag\\& = & 2 \cos \theta \sin \theta
  (m_{\Xi_c}^2 - m_{\Xi_c'}^2),\notag\\
   M_{F, 21} & =& m_8 M_{8}^{6-\bar{3}} \notag\\& = & 2 \cos \theta \sin \theta
  (m_{\Xi_c}^2 - m_{\Xi_c'}^2). 
  \label{eq:flavor-mixing-angle}
\end{eqnarray}
with the abbreviated  matrix elements as  
\begin{eqnarray}
  M_{8}^{F-I} & \equiv &  
  \langle \Xi_c^F (\vec{p}=0) |   O_8 (\vec {x}=0)
  | \Xi_c^I (\vec{p'}=0) \rangle ,
\end{eqnarray}
where $I,F = \bar{3},6$ denotes the $SU(3)_F$
representation of initial/final states. It is clear that the mixing angle can be extracted through the off diagonal terms of $M_{F}$ once the $M_{8}^{\bar{3} - 6}$ or $M_{8}^{6-\bar{3}}$ is obtained from lattice QCD and  $m_{\Xi_c}^2$ and  $m_{\Xi_c'}^2$  are experimentally determined.

Before closing this section, we wish to stress again that the masses $m_{\Xi_c^{\bar{3} / 6}}$ are eigenvalues of $H_0$
under the $SU(3)_F$ symmetry while the $m_{\Xi_c} / m_{\Xi_c'}$ are the physical masses of $\Xi_c$ and $\Xi_c'$. 

\subsection{Lattice QCD calculation of matrix elements}

In the lattice QCD, the masses $m_{\Xi_c^{\bar 3, 6}}$ can be extracted from the two-point functions (2pts) with $u s c$-type interpolators, in which the 2pts are defined as
\begin{eqnarray}
  C_2^{\bar{3}/6} (t)=
    \int d^3\vec{y}  
  T'_{\gamma' \gamma}  \langle   O_{\gamma, SU(3)}^{\bar{3}/6} (\vec{y}, t)
  \bar{O}_{\gamma', SU(3)}^{\bar{3}/6} (\vec{0}, 0) \rangle. 
\end{eqnarray} 
Here $\gamma$ and $\gamma'$ are spinor indices and $T'$ is a projection matrix.   
The interpolating  operators for the antitriplet and sextet baryons are used as~\cite{Grozin:1992td}
\begin{eqnarray}
 O_{ SU(3)}^{\bar{3}} & = & \epsilon^{abc} (q^{T a} C
  \gamma_5 s^{ b}) P_+ c^c,\\
 O_{SU(3)}^{6} & = & \epsilon^{abc} (q^{T a} C \vec{\gamma}
  s^{ b}) \cdot \vec{\gamma} \gamma_5 P_+ c^c .
\end{eqnarray}
It should be noticed that in the above definition, we have used the charm quark field defined in QCD, not in HQET. This will not affect the flavor SU(3) symmetry.

Inserting the hadronic states,  keeping the lowest two hadrons, and choosing $T'=I$, one has
\begin{eqnarray}
  C_2^{\bar{3}/6} (t) 
   & = & f^2_{\Xi^{\bar{3}/6}_c} m_{\Xi_c^{\bar{3}/6}}^4 e^{- 
  m_{\Xi_c^{\bar{3}/6}} t} (1+d_i e^{-\Delta m_{\Xi_c^{\bar{3}/6}}t}), \nonumber\\
\label{eq:2pt-parameterization}
\end{eqnarray} 
where $f_{\Xi^{\bar{3}/6}_c}$ denotes the   decay constant  of  $\Xi^{\bar{3}}_c$ or $\Xi^{6}_c$ as
\begin{eqnarray}
  \langle \vec{k} | \bar{O}_{ SU(3)}^{\bar{3}/6} (0, 0) | 0
  \rangle & = &  f_{\Xi_c^{\bar{3}/6}} m_{\Xi_c^{\bar{3}/6}}^2 
  \bar{u}(\vec{k}). 
 \end{eqnarray} 
 and $\Delta m_{\Xi_c^{\bar{3}/6}}$ describes the mass difference between the first excited states and ground states, and  $d_i$ characterizes the excited contributions to the two-point correlation. 

\begin{figure}
\centering
\includegraphics[width=0.9\linewidth]{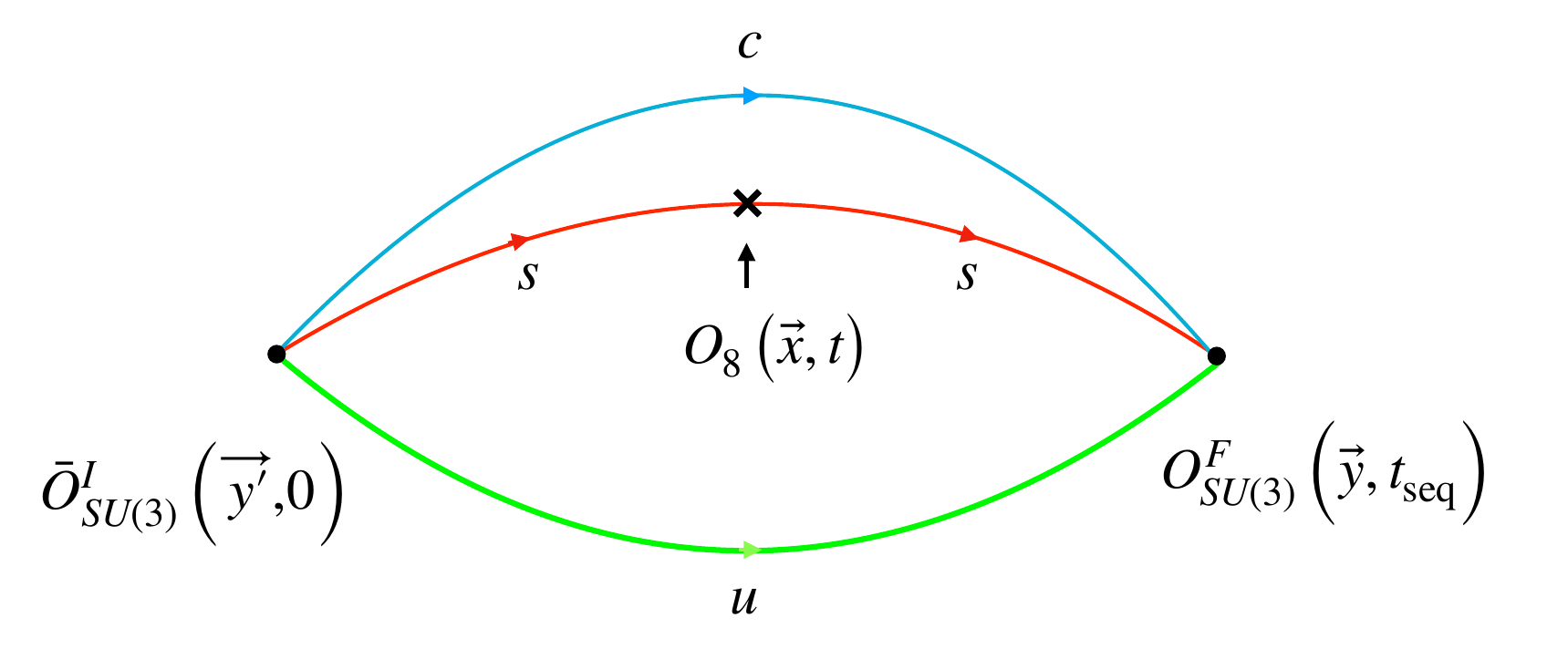}
\includegraphics[width=0.9\linewidth]{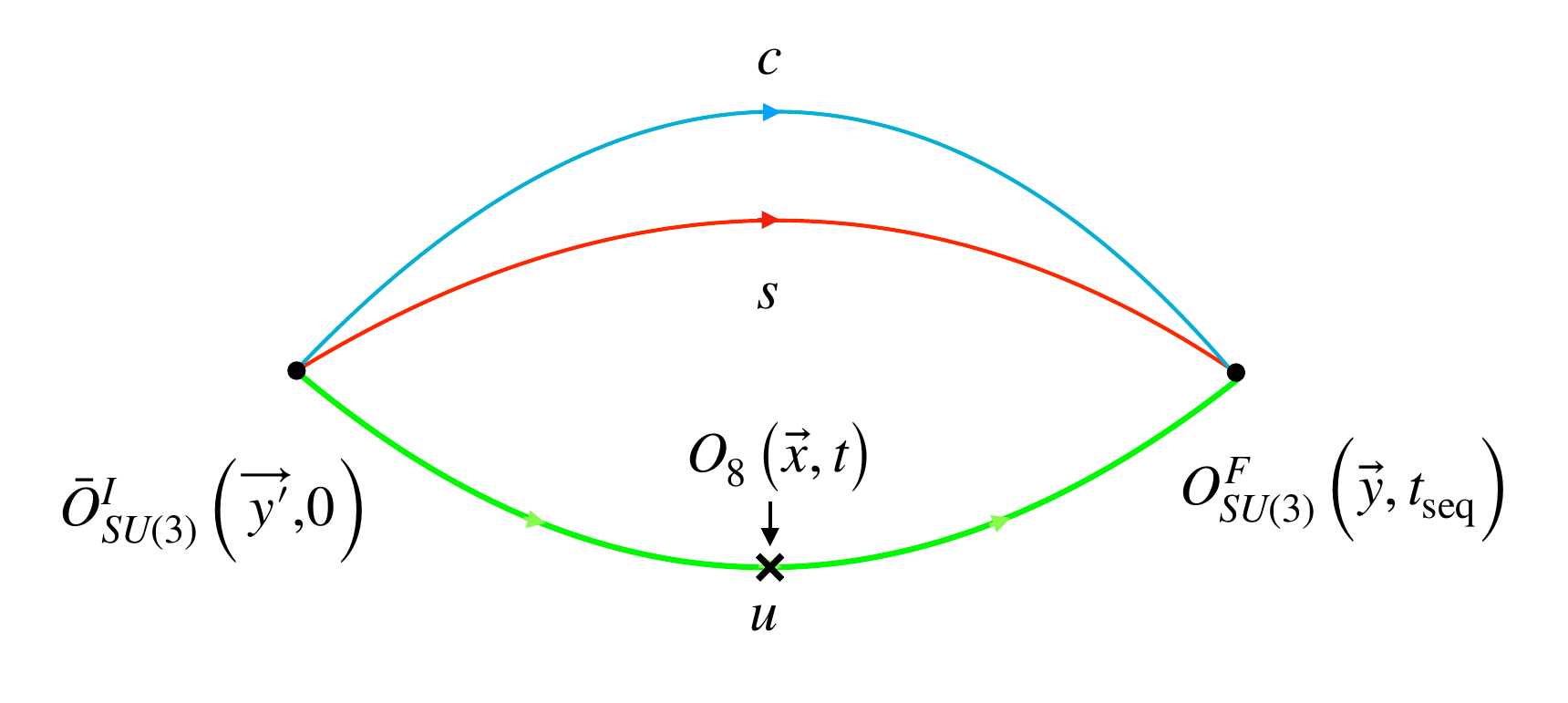}
\caption{An illustration of the off diagonal three-point correlation functions shown in Eq.(\ref{eq:three-point functions}) on the lattice. }
    \label{fig:3-point_correlation_function}
\end{figure}

The  $M_{8}^{F-I}$ can be extracted through the analysis of the three-point function (3pt) as
\begin{widetext}
\begin{align}
  C_3^{F-I} (t_{\rm seq}, t) = & \int \frac{d^3\vec{q}}{(2\pi)^3} \int d^3 \vec{y} d^3 \vec{y'} d^3 \vec{x}    e^{i \vec{q}
  \cdot \vec{x}} T_{\gamma' \gamma}  \left\langle O_{\gamma,SU(3)}^F (\vec{y},
  t_{\rm seq}) O_8 (\vec{x}, t) \bar{O}_{\gamma',SU(3)}^I (\vec{y'}, 0)
  \right\rangle,
\label{eq:three-point functions}
\end{align}
\end{widetext}
where we choose $T_{\gamma' \gamma}$ as the identity matrix to simplify the expressions and the superscript $F/I$ mean the final state and the initial state which can be $\Xi^{\bar{3}}_c/\Xi^6_c$. The momentum transfer $\vec{q}=0$ comes from the conservation of momentum of the rest initial and final state.
An illustration of the three-point correlation function is shown in Fig.~\ref{fig:3-point_correlation_function}.

It should be pointed out that for a complete analysis the flavor symmetry-breaking effects should be incorporated in both valence and sea quarks.  Contributions from sea quarks can occur through the so-called disconnected diagrams. The computation of these diagrams requires the quark propagators at all points on the lattice (the so-called all-to-all propagators), which are costly in lattice simulations. However fortunately in the current decomposition of SU(3) symmetry breaking Hamiltonian, the contribution of the disconnected diagrams which are proportional to the trace of the $O_8$ operator vanishes at the leading order. Thus our analysis is limited to the valence quark.

By inserting a complete set of eigenstates of the Hamiltonian $H_0$ between the operators,  we can simplify  Eq.~\eqref{eq:three-point functions} as
\begin{widetext}
\begin{eqnarray}
    C_3^{F-I} (t_{\rm seq}, t) & = & \frac{ M_{8}^{F-I}}{\sqrt{4m_{\Xi^I_c}m_{\Xi^F_c}}}f_{\Xi^I_c}f_{\Xi^F_c}m^2_{\Xi^I_c}m^2_{\Xi^F_c}e^{-\left(m_{\Xi^I_c}-m_{\Xi^F_c}\right)t}e^{-m_{\Xi^F_c} t_{\rm seq}}  \left(1+c_1e^{-\Delta m_{\Xi^I_c} t}\right)\left(1+c_2e^{-\Delta m_{\Xi^F_c}(t_{\rm seq}-t)}\right),\notag\\
\label{eq: 3pt-parameterization}
\end{eqnarray}
\end{widetext}
where $m_{\Xi^I_c}$ and $m_{\Xi^F_c}$ are the ground-state energies of $\Xi^{\bar{3}}_c$ and $\Xi^6_c$ and $c_i$ are parameters decoding the  excited state contamination. $\Delta m_{\Xi^I_c}$ and $\Delta m_{\Xi^F_c}$ describe the mass differences between the first excited states and ground states. 

Combining the 3pt and 2pt, one can remove the dependence on nonperturbative 
decay constants.  However, there is a remnant ambiguity in determining the sign of the $M_{8}^{F-I}$. From Eq.~\eqref{eq:2pt-parameterization}, one can notice that the two-point correlation contains the square of the decay constant, while the three-point function in Eq.~\eqref{eq: 3pt-parameterization} is proportional to the decay constant for the initial state and final state. Thus if the initial state and final states are different,  the determination of $M_{8}^{F-I}$ and accordingly the $\theta$ has a sign problem from the 3pt. 

Keeping in mind this ambiguity, one can make use of  the following ratio to suppress the contributions from the excited states: 
\begin{eqnarray}
  R & = & \sqrt{\frac{C_3^{F I} (t_{\rm seq},t ) C_3^{F I} (t_{\rm seq},t_{\rm seq}-t )}{C_2^I (t_{\rm seq})
  C_2^F (t_{\rm seq})}}. 
\label{eq: R-ratio-define}
\end{eqnarray}
Combing Eqs.\eqref{eq:2pt-parameterization} and \eqref{eq: 3pt-parameterization}, $R$ can be parametrized as
\begin{widetext}
\begin{eqnarray}
R   &=& \frac{\left|  M_{8}^{F-I}\right|}{2\sqrt{ m_{\Xi^I_c}m_{\Xi^F_c}}} \bigg(\frac{(1+c_1e^{-\Delta m_{\Xi^I_c} t})(1+c_1e^{-\Delta m_{\Xi^I_c} (t_{\rm seq}-t)})(1+c_2e^{-\Delta m_{\Xi^F_c}t})(1+c_2e^{-\Delta m_{\Xi^F_c}(t_{\rm seq}-t)})}{(1+d_1e^{-\Delta  m_{\Xi^F_c}  t_{\rm seq}})(1+d_2e^{-\Delta  m_{\Xi^I_c}  t_{\rm seq}})}\bigg)^{1/2} \notag\\
&\simeq& \frac{\left|  M_{8}^{F-I}\right|}{2\sqrt{ m_{\Xi^I_c}m_{\Xi^F_c}}} \left( \frac{(1 + c_1 e^{-\Delta m_{\Xi^I_c} t} +
  c_2 e^{-\Delta m_{\Xi^F_c}(t_{\rm seq} - t)}) (1 + c_1 e^{-\Delta m_{\Xi^I_c}  (t_{\rm seq} - t)} + c_2
  e^{-\Delta m_{\Xi^F_c} t})}{(1 + d_1 e^{-\Delta m_{\Xi^I_c} t_{\rm seq}})(1 +
  d_2 e^{-\Delta m_{\Xi^F_c} t_{\rm seq}})}  \right)^{1/2},
\label{eq: parameterization}
\end{eqnarray}
\end{widetext}
where the nonperturbative decay constants have been eliminated and temporal dependence of $R$ becomes symmetric under $t\leftrightarrow(t_{\rm seq} - t)$, which allows one to extract the values of $\left|  M_{8}^{F-I}\right|$ conveniently.

In practice, we adopt the initial state $I=\bar{3}$ and final state $F=6$ to generate the correlation functions related to the off diagonal term of $M_F$, and then extract the $|M_{8}^{6-\bar 3}|$ numerically. Based on Eq.(\ref{eq:flavor-mixing-angle}), the mixing angle can be evaluate from the formula
\begin{align}
\sin{2\theta}=\pm \frac{m_8 M_{8}^{6-\bar 3}}{m^2_{\Xi'_c}-m^2_{\Xi_c}},
\end{align}
where the $\pm$ reveals the sign ambiguity from 3pt, and cannot be uniquely fixed for the time being.

\section{Numerical Results}
\label{sec:numerics}

As shown in the previous section, one can determine the mixing angle by calculating the five quantities in Eq.~\eqref{eq:mixing_matrix}. In addition, one can also make use of $m_{\Xi_c}$ and $m_{\Xi_c'}$ and obtain the mixing angle through the simulation of the off diagonal matrix element. In the following estimate, we will adopt the latter strategy for an illustration. 

Our numerical calculations are based on the lattice QCD calculations with the gauge configurations generated by the Chinese Lattice QCD (CLQCD) Collaboration with $N_f=2+1$ flavor stout smeared clover fermions and Symanzik gauge action \cite{Hu:2023jet}. These configurations have been applied to explore different physical quantities as in Refs.~\cite{Zhang:2021oja,Wang:2021vqy,Liu:2022gxf,Xing:2022ijm}.   

For the estimation of the off diagonal matrix element, we choose one set of lattice ensembles with the lattice spacing $a = 0.108 \rm fm$.  The detailed parameters of the ensemble are listed in Table \ref{tab: configurations}. The bare charm quark mass is tuned to accommodate the spin-average value of the $J/\psi$ and $\eta_c$ masses, more details can be found in Ref.~\cite{Liu:2023feb}. 
The quark propagators are computed using the Coulomb gauge fixed wall source at one source time slice.  By choosing different reference time slices, we perform $432 \times 6$ measurements on $\rm C11P29S$ ensemble.

\begin{widetext}

\begin{table}[h!]
\renewcommand{\arraystretch}{2.0}
\setlength{\tabcolsep}{1.8mm}
\begin{tabular}{c c c c c c c c c}
\hline\hline
   Ensemble &  $\beta$ & $L^3\times T$  & $a$ (\rm fm) & $m_l^{\mathrm{b}}$  & $m_s^{\mathrm{b}}$ & $m_c^{\mathrm{b}}$ & $m_{\pi}(\rm MeV)$   & $N_{\mathrm{meas}}$    \\\hline
   C11P29S &  $6.20$       &  $24^3\times72$   &   $0.108$           &  $-0.2770$ & $-0.2315$ & $0.4780$ & $284$ & $432\times6$ \\\hline
\end{tabular}
\caption{Parameters of the ensembles used in this work, including the gauge coupling $\beta=10/g^2$, spatial lattice size $L$ and temporal $T$, lattice spacing $a$, bare quark masses $m_{l,s,c}^{\mathrm{b}}$, pion mass $m_{\pi}$ and total measurements $N_{\mathrm{meas}}$. The total measurements are equal to the number of gauge configurations times the measurements from different time slices on one configuration. } 
\label{tab: configurations}
\end{table}
\end{widetext}

\begin{figure}
\centering
\includegraphics[width=0.9\textwidth]{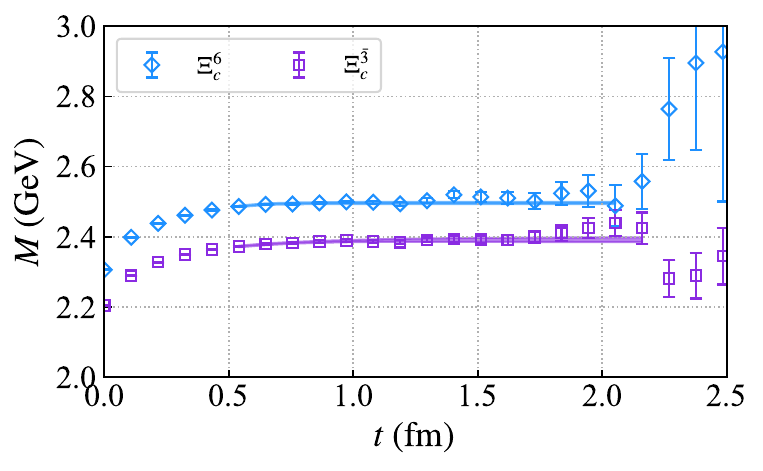}
    \caption{Effective mass of $\Xi^{\bar{3}}$ and $\Xi^6_c$ on the C11P29S ensemble. The purple markers and the corresponding fit line represent the effective mass of $\Xi^{\bar{3}}$. The blue markers denote the effective mass of $\Xi^6_c$. }
    \label{fig:C11P29S_2pt}
\end{figure}

The masses of $\Xi^{\bar{3}}$ and $\Xi^6_c$ states are extracted from fitting the 2pt via the two-state parametrization in Eq.~\eqref{eq:2pt-parameterization}, and the corresponding results are shown in Fig.~\ref{fig:C11P29S_2pt}. 
Choosing the proper time slices range, we obtain good fits with $\chi^2/\mathrm{d.o.f}=0.49$ and $\chi^2/\mathrm{d.o.f}=1.1$, and obtain $m_{\Xi^{\bar{3}}_c}=(2.395\pm 0.007)\rm GeV$ and $m_{\Xi^{\bar{3}}_c}=(2.500\pm 0.003) \rm GeV$. 

\begin{figure}
\centering
\includegraphics[width=1\textwidth]{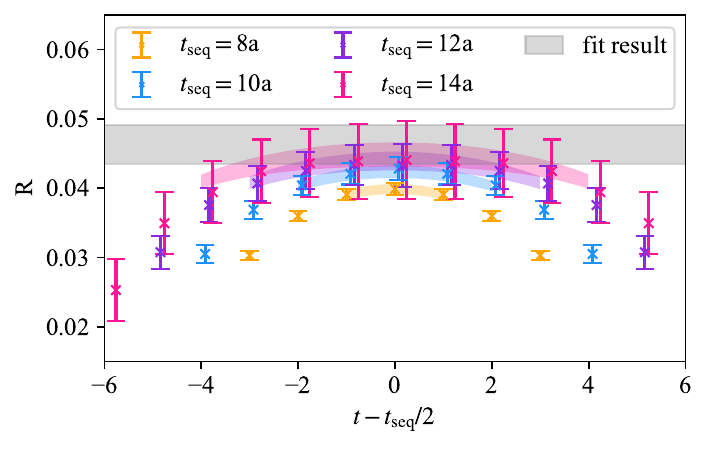}
    \caption{Joint fit of ratio $R$ as function of $t$, with $t_{\mathrm{seq}}=8a\sim14a$. In this figure, the colored bands correspond to the fitted results at each $t_{\mathrm{seq}}$, and the gray band denotes the fit results of $|M_8^{6-\bar{3}}|/(2\sqrt{ m_{\Xi^{\bar{3}}_c}m_{\Xi^6_c}})$. The $\chi^2/\rm {d.o.f}$ of this fit is about $0.17$.  }
    \label{fig:C11P29S_f}
\end{figure}

We numerically simulate the three-point function $C_3^{6-\bar 3} (t_{\rm seq}, t)$, and  adopt the parametrization  in Eq.~\eqref{eq: parameterization} to extract the matrix elements $|  M_{8}^{6-\bar{3}}|$ and  $|  M_{\bar{s} s}^{6-\bar{3}}|$, the fit result is shown in Fig.~\ref{fig:C11P29S_f}. 
To determine the mixing angle,  we quote the masses $m_{\Xi_c}=2.468$GeV and $m_{\Xi'_c}=2.578$ from Particle Data Group (PDG)~\cite{ParticleDataGroup:2022pth}. For the quark masses, their results depend on the scale, which should be compensated by the renormalization scale dependence of the  $O_8$ (or $\bar ss$) operator. Since the aim of this paper is to demonstrate the improved method used in this work, we take two values for the quark masses and include their differences as a systematic uncertainty, which in principle could be removed by a more sophisticated analysis on the lattice.   PDG gives $m_s-m_u\simeq 0.090 \rm GeV$ at $\mu=2$ GeV, and the running effects from $2$GeV to $1$GeV approximately gives a factor 1.35~\cite{ParticleDataGroup:2022pth}. So we adopt $m_s-m_u\simeq 0.12\rm GeV$ at $\mu=1$GeV in our calculation, and take into account the scale uncertainty to estimate the systematic error from quark masses. The numerical results of the matrix elements $|  M_{\bar{s} s/8}^{6-\bar{3}}|$ and mixing angle $\theta$ are collected in Tab.~\ref{tab:theta}. 

\begin{table}[h!]
\renewcommand{\arraystretch}{2.0}
\setlength{\tabcolsep}{1.8mm}
    \centering
    \begin{tabular}{cccc}
         \hline\hline
         Ensemble & $|  M_{\bar{s} s}^{6-\bar{3}}|$& $|  M_{8}^{6-\bar{3}}|$  & $|\theta|$ \\
         \hline
         $\rm C11P29S$& {$0.131(8)$} & {$0.227(14)$}    & $(0.810\pm 0.050\pm 0.200)^{\circ}$\\
         \hline
    \end{tabular}
    \caption{Results of the matrix elements $|  M_{8}^{6-\bar{3}}|$ and  $|  M_{\bar{s} s}^{6-\bar{3}}|$ (in unit of GeV), as well as the mixing angle $\theta$. The former only contains statistical uncertainty, while the latter one include both statistical and systematic uncertainties.  }
    \label{tab:theta}
\end{table}

A few remarks are given in order. 

\begin{itemize}
\item It is necessary to point out that the lattice renormalization of the 3pt and the scale dependence in quark masses are not systematically taken into account in above discussion.

\item In this calculation we only adopt one ensemble of CLQCD configurations. A near-term calculation~\cite{Liu:2023feb}, which has systematically considered the effects from physical mass extrapolation and continuum extrapolation, indicates that a result obtained from C11P29S approximately exists with a 20\% deviation from the physical one. It can also happened in the current calculation.   

\item Despite the undetermined sign,  the absolute  value for $\theta$ indicates that it is insufficient to account for the large SU(3) symmetry breaking effects in semileptonic weak decays of charmed baryons~\cite{BESIII:2015ysy,BESIII:2023vfi,Belle:2021crz,Belle:2021dgc}, and leaves the large SU(3) symmetry breaking problem unresolved. 

\item Numerical  results show that the three-point function $C_3^{6-\bar 3} (t_{\rm seq}, t)$ is negative. From Eq.~\eqref{eq: 3pt-parameterization}, one can see that if the decay constants for $\Xi_c^{\bar 3}$ and $\Xi_c^{6}$ have the same sign, the obtained mixing angle will be positive.

\item One can calculate the diagonal matrix element of the Hamiltonian, namely $M_{F,11}$ and $M_{F,22}$, which does not contain the sign ambiguity in the determination of  $M_{8}^{F-I}$. However from Eq.~\eqref{eq:flavor-mixing-angle}, one can see that the square of cosine and sine of $\theta$  appears in the relation and thus, still can not be uniquely determined. 

\end{itemize}

\section{The Mixing Angle and Heavy quark symmetry breaking}
\label{sec:HQSV}

In heavy quark effective theory,  the classification of heavy baryonic states is based on heavy quark symmetry for the heavy quark and flavor SU(3)  symmetry for the light quarks. 
In heavy quark limit,  the corresponding Lagrangian for  a heavy quark  is given as:
\begin{eqnarray}
    {\cal L}_Q = \bar h_v (iv\cdot D)h_v, 
\end{eqnarray}
where $h_v$ is the heavy quark field and $v$ denotes the velocity. In this Lagrangian, the heavy quark such as a charm quark serves as a static color source and the interaction term does not modify the spin. As a result the heavy quark  decouples with the light-quark system, and thereby charmed baryons can be classified according to the quantum number of the light-quark system. Furthermore, when light quarks in QCD Lagrangian have the same masses, the light-quark system in a charmed baryon forms an SU(3) triplet and sextet. This is how charmed baryons are classified.

In reality, the $\Xi_c$ in the triplet and $\Xi_c'$ in the sextet can mix with each other, and  this mixing requires the breaking of both heavy quark and flavor SU(3) symmetries. It is evident that only when the flavor SU(3) symmetry is broken,  baryons in different multi-plets can get entangled with each other. The requirement for breaking heavy quark symmetry can be understood as follows. In heavy quark limit, the heavy quark acts as a color source and the interacting gluon does not change the spin. Thereby light-quark systems in charmed baryons have conserved total spin and behave like a $\pi$ and $\rho$ meson with different angular momenta, respectively. If the heavy quark symmetry is not spoiled,  no source is provided to modify the spin of the light quark system, and accordingly  $\Xi_c$ and $\Xi_c'$ baryons will not mix with each other. 
It is anticipated that the mixing is proportional to $1/m_Q$.

While the constructions of baryonic states are established under both heavy quark and flavor SU(3) symmetry,  in our lattice simulation of the correlation function, we have used a finite mass for the charm quark. This explicitly breaks the heavy quark symmetry and can induce the $\Xi_c$ and $\Xi_c'$ mixing.

It is necessary to stress that in the lattice QCD simulation the charm quark can not be too large. This is due to the fact that the discretization effects are likely proportional to $m_c^2 a^2$. On the other hand, the charm quark quark can not be chosen too small. In our calculation of the mixing angle, we have firstly constructed the SU(3) symmetric hadron state and then calculated the matrix elements of symmetry breaking Hamiltonian. This is based on the spirit of perturbation theory with the expansion parameter $(m_s-m_u)/m_c$. If the charm quark mass is small the expansion parameter would be large and the perturbation could in general fail.

To investigate the heavy quark mass dependence of the mixing angle, and to predict the behavior at the heavy quark limit, we vary the charm quark mass $m_c$ in lattice calculation, and more explicitly, we have chosen $m_{\Xi^{\bar{3}}_c}=2.047(6), 2.183(6), 2.309(6), 2.401(6), 2.535(6), 2.637(6),\\ 
2.734(6)\rm GeV $. 
The mixing angle is correspondingly extracted and the results are shown in Tab.~\ref{tab:theta_mQ}.

\begin{table}[h!]
\renewcommand{\arraystretch}{2.0}
\setlength{\tabcolsep}{1.8mm}
    \centering
    \begin{tabular}{ccccc}
         \hline\hline  
        $m_{\Xi^{\bar 3}_c}(\rm GeV)$ & {$2.047(6)$} & {$2.183(6)$}    & $2.309(6)$& {$2.401(6)$} \\
        $|\theta| (^\circ)  $ &$1.03(6)$ &$0.94(5)$ & $0.86(5)$ & $0.81(5)$  \\
        $m_{\Xi^{\bar 3}_c}(\rm GeV)$& {$2.535(6)$}    & $2.637(6)$& $2.734(6)$&\\
        $|\theta| (^\circ)$& $0.75(5)$ &$0.71(5)$ & $0.67(5)$&\\
         \hline\hline  
    \end{tabular}
    \caption{Results of the mixing angle and the dependence on heavy baryon mass.  Only statistical results are included in the results.   }
    \label{tab:theta_mQ}
\end{table}

\begin{figure}
\centering
\includegraphics[width=1\linewidth]{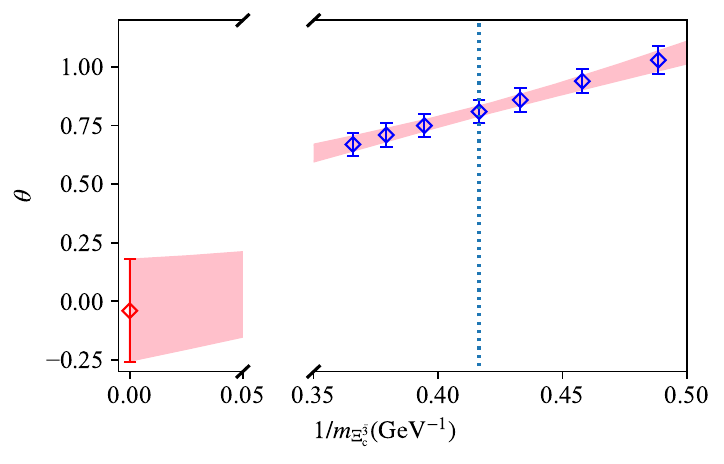}
\caption{The heavy quark mass dependence of mixing angle $\theta$. The blue data points denote the results calculated from different charm quark masses, and the dashed line denotes the physical one. The red band shows the fit result of $\theta$ as a function of $1/m_{\Xi^{\bar{3}}_c}$ based on Eq.(\ref{eq:mcdependenceoftheta}), and the red data point indicates the value of $\theta$ is consistent to 0 at $m_c$ tends to infinity.  }
    \label{fig:mc_depend}
\end{figure}

From this table one can see that the mixing angle will decrease with the increase of charm quark and charmed baryon mass. 
We then employ a  fit ansatz for the mixing angle $\theta$ as a function of $m_{\Xi^{\bar{3}}_c}$
\begin{align}
	\theta = \frac{c_1}{m_{\Xi^{\bar{3}}_c}} + \frac{c_2}{m^2_{\Xi^{\bar{3}}_c}}+c_3, \label{eq:mcdependenceoftheta}
\end{align}
with fit results $c_1=1.25(85)$GeV, $c_2=1.9(1.1)\mathrm{GeV^2}$ and $c_3=-0.04(22)$. The results of $\theta$ extracted from different $m_c$ as well as the fit band are illustrated in Fig.~\ref{fig:mc_depend}.  It should be highlighted that from the fit result one can see that the mixing angle is consistent with 0 when the charm quark mass tends to infinity, shown as the red data point in the figure. It demonstrates the mixing effect vanishes in the heavy quark limit. This validates the classification of baryons in the heavy quark limit.

\section{Summary} 

In this work, we have developed an improved method to explore the  $\Xi_c- \Xi_c'$ mixing which arises from the flavor SU(3) and heavy quark symmetry breaking effects.  The recipe in this method is summarized as follows. 
\begin{itemize}
\item First, the flavor eigenstates are constructed under the flavor SU(3)symmetry. The corresponding masses can be determined via an explicit nonperturbative calculation using lattice QCD simulation or QCD sum rules.  
\item The SU(3) symmetry breaking contributions are treated as perturbative corrections. Matrix elements of the mass operators which break the flavor SU(3) symmetry sandwiched by the flavor eigenstates are then calculated.
\item Diagonalizing the corresponding matrix of Hamiltonian gives the mass eigenstates of the full Hamiltonian and determines the corresponding mixing. 
\item  Using the physical masses from data,  one can actually determine the mixing angle by only calculating the off diagonal matrix elements.
\end{itemize} 

Estimating an off diagonal matrix element, we have extracted the mixing angle between the $\Xi_c$ and $\Xi_c'$, with a sign ambiguity. Preliminary numerical results for the mixing angle confirm the previous observation that such mixing is not able to explain the large SU(3) symmetry breaking in semileptonic charmed baryon decays. 

It should be pointed out that in this method only the leading order contributions from the symmetry breaking terms are taken into account, and it is based on a perturbative expansion in terms of $(m_s-m_u)/\Lambda$ with $\Lambda$ being the hadronic scale. In the $\Xi_c-\Xi_c'$ mixing the heavy quark symmetry also needs to be broken, introducing a factor $\Lambda/m_c$. Other interesting examples such as the $K_1(1270)$ and $K_1(1400)$ mixing also due to the flavor SU(3) symmetry breaking can be analyzed similarly. 

Though in our illustration, the lattice QCD has been used to calculate the matrix element, this method can be applied with other nonperturbative approaches like the QCD sum rules~\cite{Sun:2023noo}.  Following this spirit, a recent analysis~\cite{Deng:2023qaf} has estimated the QED contribution to $\Xi_c^+-\Xi_c^{\prime+}$ mixing angle. 

\section*{Acknowledgements}
We thank Liuming Liu, Peng Sun, Wei Sun, Jin-Xin Tan,  Yi-Bo Yang for the collaboration on Ref.~\cite{Liu:2023feb} and valuable  discussions, and CLQCD for providing the lattice ensembles. 
W. Wang would like to thank Feng-Kun Guo, Jia-Jun Wu, and Qiang Zhao for inspiring discussions. 
This work is supported in part by Natural Science Foundation of China under grant No.U2032102, 12125503, 12061131006, 12335003 and 12375069.  The computations in this paper were run on the Siyuan-1 cluster supported by the Center for High Performance Computing at Shanghai Jiao Tong University, and Advanced Computing East China Sub-center. The LQCD calculations were performed using the Chroma software suite~\cite{Edwards:2004sx} and QUDA~\cite{Clark:2009wm,Babich:2011np,Clark:2016rdz} through HIP programming model~\cite{Bi:2020wpt}.

\begin{appendix}

\section{Another decomposition of symmetry breaking Hamiltonian}

In addition to the decomposition of Hamiltonian used in the main text that is based on a complete  SU(3) symmetry analysis, one can also adopt another equivalent way.  where the symmetry breaking term comes from the deviation between $u/d$ and $s$ quark masses:
\begin{eqnarray}
\Delta  \mathcal{L} & = & - \bar{s} (m_s - m_{u}) s.
\end{eqnarray}
The pertinent  Hamiltonian is correspondingly derived as
\begin{align}
 H &= \int d^3 \vec{x}  \left[\frac{\partial\mathcal{L}}{\partial\dot{\psi}(\vec{x})}\dot{\psi}(\vec{x})+\frac{\partial\mathcal{L}}{\partial\dot{\bar{\psi}}(\vec{x})}\dot{\bar{\psi}}(\vec{x})-\mathcal{L}\right] \nonumber\\
 &\equiv H_0 + \Delta H,  
\end{align}
with 
\begin{eqnarray}
\Delta H= (m_s-m_{u} )  \int d^3 \vec{x}   \bar{s}s(\vec{x}). 
\label{eq:Hamiltonian_II}
\end{eqnarray}
Compared to Eq.~\eqref{eq:Hamiltonian_I}, one can see that there is a correspondence between the symmetry breaking Hamiltonian:
\begin{eqnarray}
    m_8\to m_s-m_u, \;\;\; O_8\to \bar ss. 
\end{eqnarray}

Without considering the disconnected diagrams, the two forms give an equivalent result at leading order in $m_s-m_u$. 
For example, neglecting higher order SU(3) symmetry breaking effects and disconnected diagrams,  the off diagonal matrix element $M_{8}^{6-\bar 3}$
can be simplified as:
\begin{eqnarray}
  M_{8}^{6-\bar 3} 
  &=& \frac{1}{\sqrt3} 
  \langle \Xi_c^6   |  \bar uu +\bar dd -2\bar ss
  | \Xi_c^{\bar 3} \rangle \nonumber\\
  &=& \frac{1}{\sqrt3}   \langle \Xi_c^6   | -3\bar ss
  | \Xi_c^{\bar 3} \rangle\nonumber\\
  &=& -\sqrt3   \langle \Xi_c^6   |\bar ss
  | \Xi_c^{\bar 3} \rangle \nonumber\\
  &\equiv& -\sqrt3  M_{\bar ss}^{6-\bar 3}. 
\end{eqnarray}
In deriving the above equation, we have made use of the fact that the anti-triplet state is anti-symmetric under the interchange of $u/d\leftrightarrow s$  and the sextet state is symmetric.

\begin{figure}
\centering
\subfigure[]{\label{fig:subfig:a}
\includegraphics[width=0.9\linewidth]{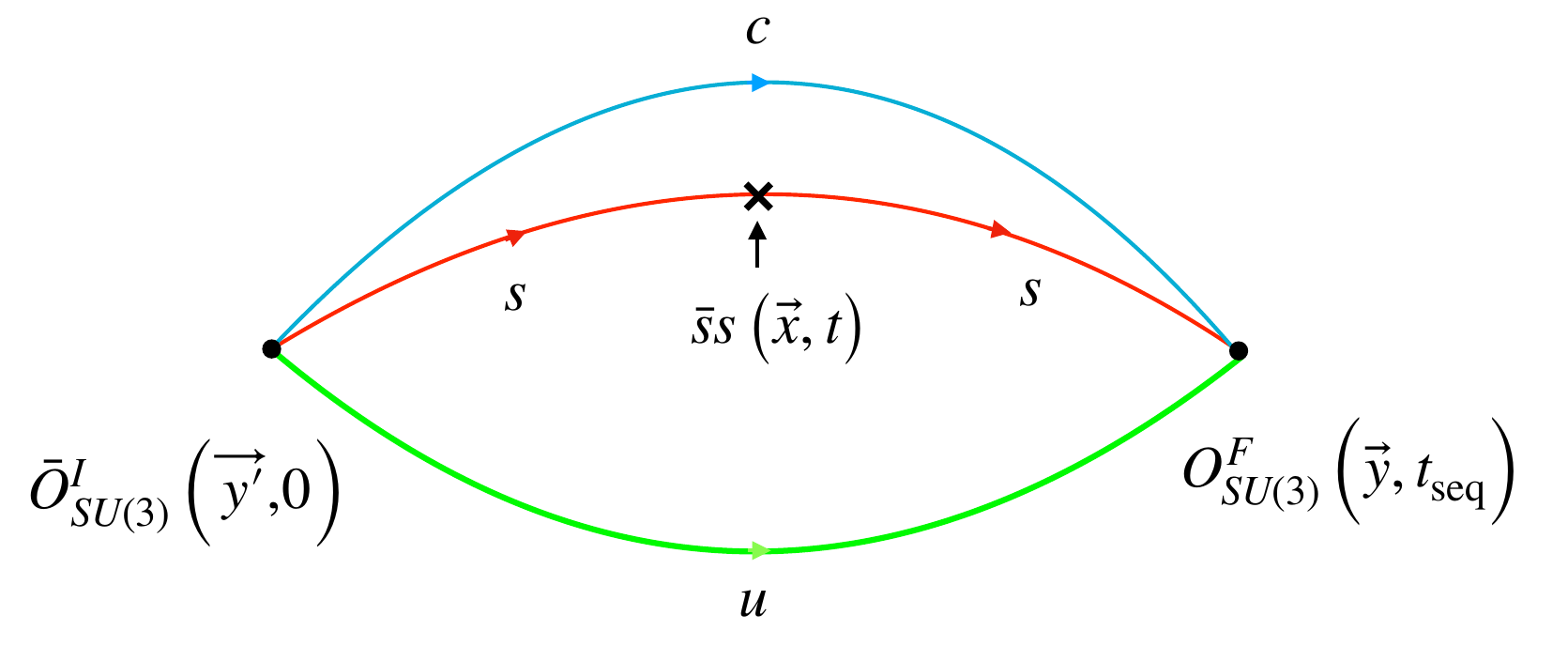}}
\vspace{0.01\linewidth}
\subfigure[]{\label{fig:subfig:b}
\includegraphics[width=0.9\linewidth]{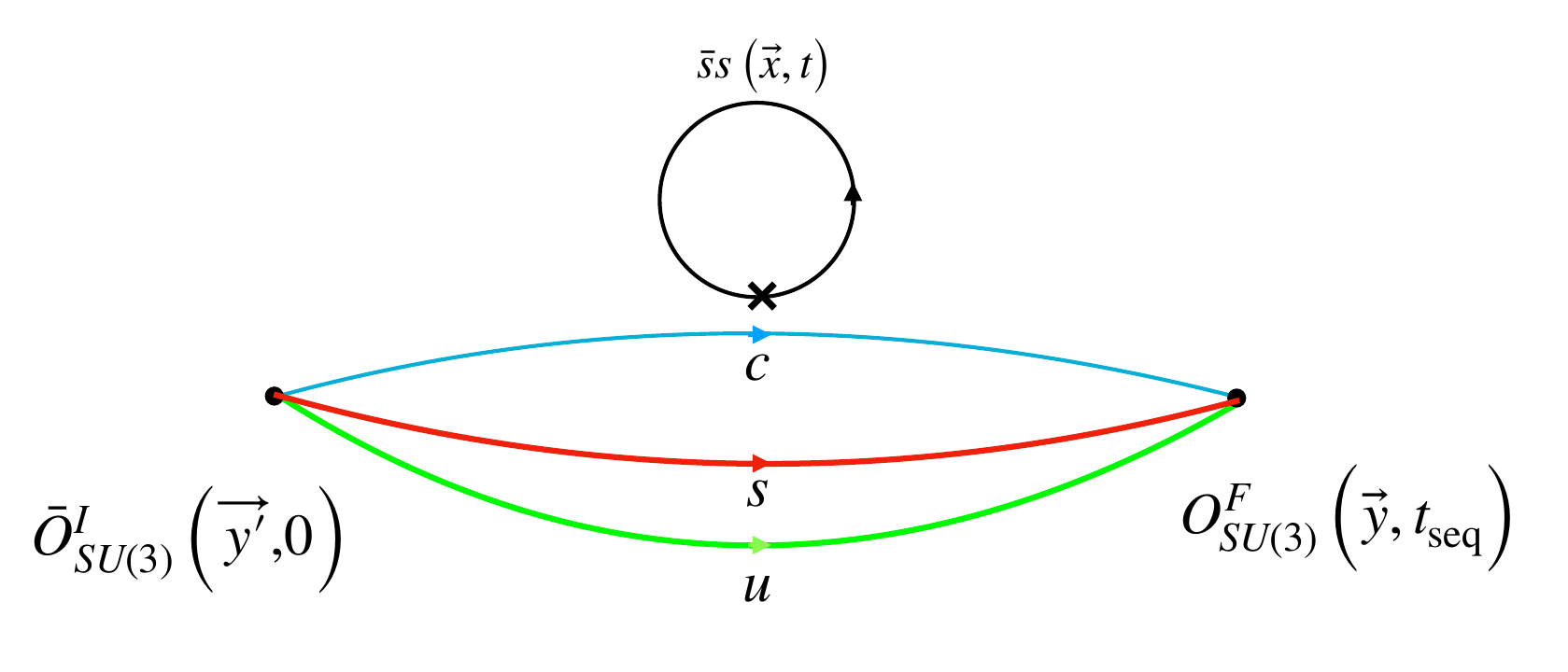}}
\caption{An illustration of the three-point correlation functions using the Hamiltonian in Eq.~\eqref{eq:Hamiltonian_II}. Compared to the decomposition in the main text, this form receives contributions from both connected diagram (a) and disconnected diagram (b). }
\label{fig:3-point_correlation_function_II}
\end{figure}

The illustration diagrams for the corresponding  3pt  are shown in Fig.~\ref{fig:3-point_correlation_function_II}. Unlike the results in the main text with the decomposition in Eq.~\eqref{eq:Hamiltonian_I},  the correlation function under this decomposition  receives contributions from disconnected diagrams as shown in panel (b), which are difficult to evaluate.

\end{appendix}


\begin{thebibliography}{}
\bibitem{BESIII:2015ysy}
M.~Ablikim \textit{et al.} [BESIII],
Phys. Rev. Lett. \textbf{115}, no.22, 221805 (2015)
doi:10.1103/PhysRevLett.115.221805
[arXiv:1510.02610 [hep-ex]].

\bibitem{BESIII:2023vfi}
M.~Ablikim \textit{et al.} [BESIII],
[arXiv:2306.02624 [hep-ex]].

\bibitem{Belle:2021crz}
Y.~B.~Li \textit{et al.} [Belle],
Phys. Rev. Lett. \textbf{127}, no.12, 121803 (2021)
doi:10.1103/PhysRevLett.127.121803
[arXiv:2103.06496 [hep-ex]].

\bibitem{Belle:2021dgc}
Y.~B.~Li \textit{et al.} [Belle],
Phys. Rev. D \textbf{105}, no.9, L091101 (2022)
doi:10.1103/PhysRevD.105.L091101
[arXiv:2112.10367 [hep-ex]].

\bibitem{Lu:2016ogy}
C.~D.~L\"u, W.~Wang and F.~S.~Yu,
Phys. Rev. D \textbf{93}, no.5, 056008 (2016)
doi:10.1103/PhysRevD.93.056008
[arXiv:1601.04241 [hep-ph]].

\bibitem{He:2018php}
X.~G.~He and W.~Wang,
Chin. Phys. C \textbf{42}, no.10, 103108 (2018)
doi:10.1088/1674-1137/42/10/103108
[arXiv:1803.04227 [hep-ph]].

\bibitem{He:2018joe}
X.~G.~He, Y.~J.~Shi and W.~Wang,
Eur. Phys. J. C \textbf{80}, no.5, 359 (2020)
doi:10.1140/epjc/s10052-020-7862-5
[arXiv:1811.03480 [hep-ph]].

\bibitem{Wang:2018utj}
W.~Wang and J.~Xu,
Phys. Rev. D \textbf{97}, no.9, 093007 (2018)
doi:10.1103/PhysRevD.97.093007
[arXiv:1803.01476 [hep-ph]].

\bibitem{ParticleDataGroup:2022pth}
R.~L.~Workman \textit{et al.} [Particle Data Group],
PTEP \textbf{2022}, 083C01 (2022)
doi:10.1093/ptep/ptac097

\bibitem{He:2021qnc}
X.~G.~He, F.~Huang, W.~Wang and Z.~P.~Xing,
Phys. Lett. B \textbf{823}, 136765 (2021)
doi:10.1016/j.physletb.2021.136765
[arXiv:2110.04179 [hep-ph]].

\bibitem{Geng:2022yxb}
C.~Q.~Geng, X.~N.~Jin and C.~W.~Liu,
Phys. Lett. B \textbf{838}, 137736 (2023)
doi:10.1016/j.physletb.2023.137736
[arXiv:2210.07211 [hep-ph]].

\bibitem{Franklin:1981rc}
J.~Franklin, D.~B.~Lichtenberg, W.~Namgung and D.~Carydas,
Phys. Rev. D \textbf{24}, 2910 (1981)
doi:10.1103/PhysRevD.24.2910

\bibitem{Franklin:1996ve}
J.~Franklin,
Phys. Rev. D \textbf{55}, 425-426 (1997)
doi:10.1103/PhysRevD.55.425
[arXiv:hep-ph/9606326 [hep-ph]].

\bibitem{Ito:1996mr}
T.~Ito and Y.~Matsui,
Prog. Theor. Phys. \textbf{96}, 659-664 (1996)
doi:10.1143/PTP.96.659
[arXiv:hep-ph/9605289 [hep-ph]].

\bibitem{Aliev:2010ra}
T.~M.~Aliev, A.~Ozpineci and V.~Zamiralov,
Phys. Rev. D \textbf{83}, 016008 (2011)
doi:10.1103/PhysRevD.83.016008
[arXiv:1007.0814 [hep-ph]].

\bibitem{Matsui:2020wcc}
Y.~Matsui,
Nucl. Phys. A \textbf{1008}, 122139 (2021)
doi:10.1016/j.nuclphysa.2021.122139
[arXiv:2011.09653 [hep-ph]].

\bibitem{Geng:2022xfz}
C.~Q.~Geng, X.~N.~Jin, C.~W.~Liu, X.~Yu and A.~W.~Zhou,
Phys. Lett. B \textbf{839}, 137831 (2023)
doi:10.1016/j.physletb.2023.137831
[arXiv:2212.02971 [hep-ph]].

\bibitem{Liu:2022igi}
C.~W.~Liu and C.~Q.~Geng,
Phys. Rev. D \textbf{107}, no.1, 013006 (2023)
doi:10.1103/PhysRevD.107.013006
[arXiv:2211.12960 [hep-ph]].

\bibitem{Ke:2022gxm}
H.~W.~Ke and X.~Q.~Li,
Phys. Rev. D \textbf{105}, no.9, 096011 (2022)
doi:10.1103/PhysRevD.105.096011
[arXiv:2203.10352 [hep-ph]].

\bibitem{Xing:2022phq}
Z.~P.~Xing and Y.~j.~Shi,
Phys. Rev. D \textbf{107}, no.7, 074024 (2023)
doi:10.1103/PhysRevD.107.074024
[arXiv:2212.09003 [hep-ph]].

\bibitem{Liu:2023feb}
H.~Liu, L.~Liu, P.~Sun, W.~Sun, J.~X.~Tan, W.~Wang, Y.~B.~Yang and Q.~A.~Zhang,
Phys. Lett. B \textbf{841}, 137941 (2023)
doi:10.1016/j.physletb.2023.137941
[arXiv:2303.17865 [hep-lat]].

\bibitem{Brown:2014ena}
Z.~S.~Brown, W.~Detmold, S.~Meinel and K.~Orginos,
Phys. Rev. D \textbf{90}, no.9, 094507 (2014)
doi:10.1103/PhysRevD.90.094507
[arXiv:1409.0497 [hep-lat]].

\bibitem{Sun:2023noo}
X.~Y.~Sun, F.~W.~Zhang, Y.~J.~Shi and Z.~X.~Zhao,
Eur. Phys. J. C \textbf{83}, no.10, 961 (2023)
doi:10.1140/epjc/s10052-023-12042-4
[arXiv:2305.08050 [hep-ph]].

\bibitem{Grozin:1992td}
A.~G.~Grozin and O.~I.~Yakovlev,
Phys. Lett. B \textbf{285}, 254-262 (1992)
doi:10.1016/0370-2693(92)91462-I
[arXiv:hep-ph/9908364 [hep-ph]].

\bibitem{Hu:2023jet}
Z.~C.~Hu, B.~L.~Hu, J.~H.~Wang, M.~Gong, L.~Liu, P.~Sun, W.~Sun, W.~Wang, Y.~B.~Yang and D.~J.~Zhao,
[arXiv:2310.00814 [hep-lat]].

\bibitem{Zhang:2021oja}
Q.~A.~Zhang, J.~Hua, F.~Huang, R.~Li, Y.~Li, C.~L\"u, C.~D.~Lu, P.~Sun, W.~Sun and W.~Wang, \textit{et al.}
Chin. Phys. C \textbf{46}, no.1, 011002 (2022)
doi:10.1088/1674-1137/ac2b12
[arXiv:2103.07064 [hep-lat]].

\bibitem{Wang:2021vqy}
G.~Wang \textit{et al.} [\ensuremath{\chi}QCD],
Phys. Rev. D \textbf{106}, no.1, 014512 (2022)
doi:10.1103/PhysRevD.106.014512
[arXiv:2111.09329 [hep-lat]].

\bibitem{Liu:2022gxf}
H.~Liu, J.~He, L.~Liu, P.~Sun, W.~Wang, Y.~B.~Yang and Q.~A.~Zhang,
Sci. China Phys. Mech. Astron. \textbf{67}, no.1, 211011 (2024)
doi:10.1007/s11433-023-2205-0
[arXiv:2207.00183 [hep-lat]].

\bibitem{Xing:2022ijm}
H.~Xing, J.~Liang, L.~Liu, P.~Sun and Y.~B.~Yang,
[arXiv:2210.08555 [hep-lat]].

\bibitem{Deng:2023qaf}
Z.~F.~Deng, Y.~J.~Shi, W.~Wang and J.~Zeng,
[arXiv:2309.16386 [hep-ph]].

\bibitem{Edwards:2004sx}
R.~G.~Edwards \textit{et al.} [SciDAC, LHPC and UKQCD],
Nucl. Phys. B Proc. Suppl. \textbf{140}, 832 (2005)
doi:10.1016/j.nuclphysbps.2004.11.254
[arXiv:hep-lat/0409003 [hep-lat]].

\bibitem{Clark:2009wm}
M.~A.~Clark \textit{et al.} [QUDA],
Comput. Phys. Commun. \textbf{181}, 1517-1528 (2010)
doi:10.1016/j.cpc.2010.05.002
[arXiv:0911.3191 [hep-lat]].

\bibitem{Babich:2011np}
R.~Babich \textit{et al.} [QUDA],
doi:10.1145/2063384.2063478
[arXiv:1109.2935 [hep-lat]].

\bibitem{Clark:2016rdz}
M.~A.~Clark \textit{et al.} [QUDA],
[arXiv:1612.07873 [hep-lat]].

\bibitem{Bi:2020wpt}
Y.~J.~Bi, Y.~Xiao, W.~Y.~Guo, M.~Gong, P.~Sun, S.~Xu and Y.~B.~Yang,
PoS \textbf{LATTICE2019}, 286 (2020)
doi:10.22323/1.363.0286
[arXiv:2001.05706 [hep-lat]].

\end{thebibliography}
\end{document}